# Superconductivity at 53.5 K in GdFeAsO$_{1-\delta}$

**Jie Yang , Zheng-Cai Li, Wei Lu, Wei Yi, Xiao-Li Shen,**

**Zhi-An Ren\*, Guang-Can Che, Xiao-Li Dong, Li-Ling Sun, Fang Zhou, Zhong-Xian Zhao\***

National Laboratory for Superconductivity, Institute of Physics and Beijing National Laboratory for Condensed Matter Physics, Chinese Academy of Sciences, P. O. Box 603, Beijing 100190, P. R. China

Abstract:

Here we report the fabrication and superconductivity of the iron-based arsenic-oxide GdFeAsO$_{1-\delta}$ compound with oxygen-deficiency, which has an onset resistivity transition temperature at 53.5 K. This material has a same crystal structure as the newly discovered high-$T_c$ ReFeAsO$_{1-\delta}$ family (Re = rare earth metal) and a further reduced crystal lattice, while the $T_c$ starts to decrease compared with the SmFeAsO$_{1-\delta}$ system.

Since the discovery of layered copper oxide superconductors [1], extensive researches have been performed to understand the mechanism of high temperature superconductivity and to explore higher $T_c$ materials. It is believed that strong electron correlation and layered structure may play an important role. However, the $T_c$ of the non-copper-based superconductors is still lower than 40 K by the prediction of BCS theory and the experimental facts. Recently, superconductivity in iron- and nickel-based layered quaternary compounds has been reported: LaOFeP ($T_c \sim$ 4K) [2], LaONiP ($T_c \sim$ 3K) [3], then LaOFeAs at 26 K when substitution of As for P and F-doping at oxygen site [4]. These discoveries have attracted much interest in further experiments and theoretical studies. Subsequently, the $T_c$ rapidly increases to above 50 K in the ReFeAsO$_{1-x}$F$_x$ family [5-9], with the replacement of La by other light rare-earth elements like Pr, Nd and Sm etc. Instead of F-doping, we have succeeded in synthesizing the ReFeAsO$_{1-\delta}$ superconductors recently, which have better superconducting properties [10]. These quaternary superconductors crystallize with the tetragonal layered ZrCuSiAs structure, in the space group of P4/nmm, which has a structure of alternating Fe-As layer and Re-O layer, similar to that of cuprates. As for the Re = Gd system, we have discovered its superconductivity previously in some multi-phased samples, here we report the superconductivity in GdFeAsO$_{1-\delta}$ and GdFeAsO$_{1-x}$F$_x$ systems.

A series of superconductors with nominal compositions of oxygen-deficient GdFeAsO$_{1-\delta}$ and F-doped GdFeAsO$_{1-x}$F$_x$ were prepared by the high pressure (HP) synthesis method. The starting chemicals Gd chips, As, Fe, Fe$_2$O$_3$, FeF$_2$ powders are all with the purity better than 99.99%. At the first step, GdAs powder was obtained by reacting Gd pieces and As powders at 650 °C for 12 hours and then 1150 °C for 12 hours. The starting materials were mixed together according to the nominal ratio, then ground thoroughly and pressed into small pellets. The pellets were sealed in boron nitride crucibles and sintered in the high pressure synthesis apparatus under a pressure of 6 GPa and temperature of 1350 °C for 2 hours. The HP samples are hard and can be polished to shiny mirror-surface.

The phase purity and structural identification were characterized by powder X-ray diffraction (XRD) analysis on an MXP18A-HF type diffractometer with Cu-K$_\alpha$ radiation from 20° to 80° with a step of 0.01°. The XRD results indicate that the main phase of all F-doped and oxygen-deficient samples adopts the same ReFeAsO structure with slight impurity phases. The impurity phases have been identified to be some by-products and do not superconduct at the measuring temperature. Fig.1 shows the comparison of a typical XRD pattern for a nominal GdFeAsO$_{0.85}$ sample synthesized by HP method and an ambient-pressure (AP) synthesized undoped GdFeAsO sample. The lattice parameters are a = 3.890(4) Å, c = 8.383(2) Å for the HP sample of GdFeAsO$_{0.85}$, and a = 3.903(7) Å, c = 8.453(1) Å for AP sample of GdFeAsO, which indicates a clear shrinkage for the oxygen-deficient sample compared with the undoped sample. All samples with either F-doping or oxygen-deficiency have shrunken crystal lattices. The lattice parameters of Gd-system is smaller than that of Sm-system reported in our previous paper [10], which shows a further enhanced chemical pressure on the Fe-As plane in the Gd-system.

The resistivity of all samples was measured by the standard four-probe method from 4 K to 300 K. Slight F-doping and oxygen-deficiency both lead to the occurrence of superconductivity in this system, while samples with oxygen-deficiency were found to have higher $T_c$ comparing with F-doped ones, and all samples have metallic behavior up to 300 k. The sample with a nominal composition of GdFeAsO$_{0.85}$ was found to have the highest $T_c$ in this Gd-system, while for the F-doping system, the nominal GdFeAsO$_{0.8}$F$_{0.2}$ was found to have the highest $T_c$, and the corresponding resistivity curves are shown in Fig. 2. As it can be seen in the inset that a clear superconducting onset transition ($T_c$(onset)) occurred at 53.5 K and a zero resistivity transition ($T_c$(zero)) at 52.3 K for the GdFeAsO$_{0.85}$; while for the GdFeAsO$_{0.8}$F$_{0.2}$, the $T_c$(onset) and $T_c$(zero) are at 51.2 K and 45.5 K. Comparing with SmFeAsO$_{1-\delta}$ superconductor, the $T_c$ starts to decrease, this may indicate that the $T_c$ has reached the maximum in the Sm-based system by the inner chemical pressure effect for the ReFeAsO$_{1-\delta}$ family.

The DC magnetization was measured using a Quantum Design MPMS XL-1 system. For an experiment cycle the sample was cooled to 1.8 K in zero field cooling (ZFC) and data were gathered when warming in an applied field, then cooled again under an applied field (FC) and measured when warming up. The DC-susceptibility data (measured under a magnetic field of 1 Oe) of the GdFeAsO$_{0.85}$ are shown in Fig. 3, with the differential ZFC curve at the right panel. The sharp magnetic transitions on DC curves indicate the good quality of this superconducting component. The onset diamagnetic superconducting transition temperature determined from the differential ZFC curve is 53.5 K, same as the onset transition point on the corresponding resistivity curve.

In conclusion, we have succeeded in preparing the GdFeAsO superconductors by both of F-doping and oxygen-deficiency. The oxygen-deficient samples were found to have better superconducting properties in this system and the highest $T_c$ is at 53.5 K for the nominal GdFeAsO$_{0.85}$ composition.

Acknowledgements:

We thank Mrs. Shun-Lian Jia for her kind helps in resistivity measurements. This work is supported by Natural Science Foundation of China (NSFC, No. 50571111 & 10734120) and 973 program of China (No. 2006CB601001 & 2007CB925002). We also acknowledge the support from EC under the project COMEPHS TTC.

Corresponding Author:

Zhi-An Ren: renzhian@aphy.iphy.ac.cn

Zhong-Xian Zhao: zhxzhao@aphy.iphy.ac.cn

Figure 1: The typical XRD patterns for the nominal GdFeAsO$_{0.85}$ sample synthesized by HP method (upper line) and GdFeAsO by AP method (lower line); the vertical bars indicate the calculated diffraction peaks for the GdFeAsO.

Figure 2: The temperature dependence of resistivity for the nominal composition of GdFeAsO$_{0.85}$ and GdFeAsO$_{0.8}$F$_{0.2}$ synthesized by HP method.

Figure 3: The temperature dependence of the DC-susceptibility, and differential ZFC curve for the nominal GdFeAsO$_{0.85}$ sample synthesized by HP method.

.

Figure.1

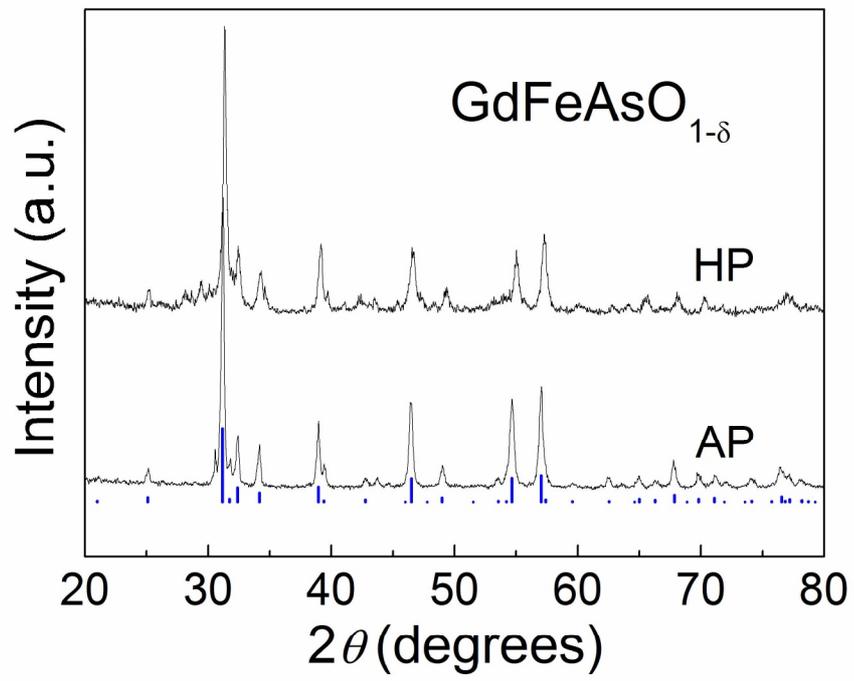

Figure.2

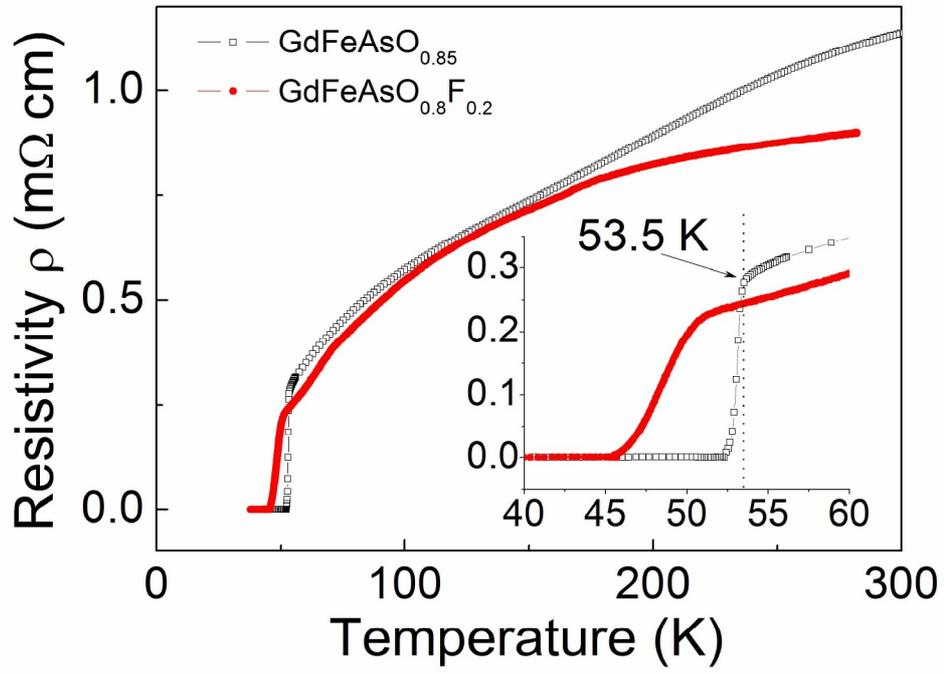

Figure.3

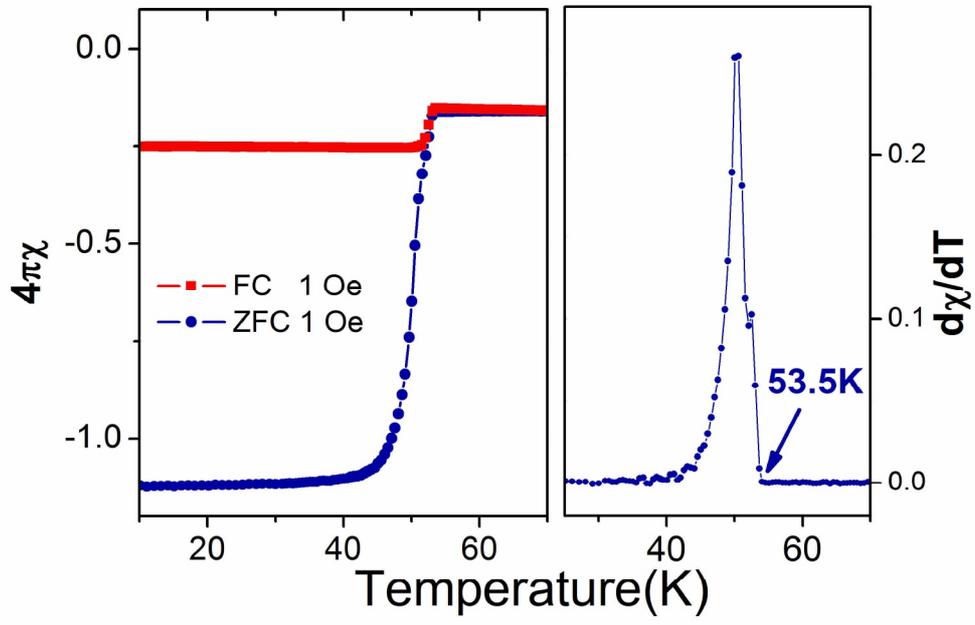